\documentstyle[PASJadd,12pt]{PASJ95}
\draft
\begin{document}

\title{Formation of Intermediate-Mass Black Holes in Circumnuclear 
       Regions of Galaxies}

\author{Yoshiaki {\sc Taniguchi}, Yasuhiro {\sc Shioya} \\
{\it Astronomical Institute, Graduate School of Science, Tohoku University,}\\
{\it Aramaki, Aoba, Sendai 980-8578} \\
{\it E-mail (YT): tani@astr.tohoku.ac.jp}
\\[6pt]
Takeshi G. {\sc Tsuru} \\
{\it Physics Department, Graduate School of Science, Kyoto University,} \\
{\it Kitashirakawa, Sakyo, Kyoto 606-8502} \\
and \\
Satoru {\sc Ikeuchi} \\
{\it Physics Department, Graduate School of Science, Nagoya University,} \\
{\it Furo, Chikusa, Nagoya 464-8602}}
%
%
\abst{
Recent high-resolution X-ray imaging studies have discovered
possible candidates of intermediate-mass black holes with masses
of $M_\bullet \sim 10^{2-4} \MO$ in circumnuclear regions
of many (disk) galaxies. It is known that a large number of 
massive stars are formed in a circumnuclear giant H {\sc ii} region.
Therefore, we propose that continual merger of compact remnants left from 
these massive stars is responsible for the formation of such an 
intermediate-mass black hole within a timescale of $\sim 10^9$ years.
A necessary condition is that several hundreds of massive
stars are formed in a compact region with a radius of a few pc.
} 
\kword{black holes --- galaxies: starburst --- 
galaxies: star clusters --- X-rays: sources}

\maketitle
\thispagestyle{headings}


\section{Introduction}

Recent sensitive, high-resolution X-ray imaging studies have found
unusually bright, compact X-ray sources in circumnuclear regions
of many (disk) galaxies [Colbert \& Mushotzky
1999 (hereafter CM99), Okada et al. 1998;
Ptak \& Griffiths 1999; Matsumoto \& Tsuru 1999; Makishima et al. 2000].
Prior to these observations, it has been known that some nearby galaxies
have luminous X-ray sources in their circumnuclear regions
(Fabbiano \& Trinchieri 1987; Kohmura et al. 1994; Petre et al. 1994;
Takano et al. 1994; Colbert et al. 1995;
Reynolds et al. 1997; see for a review Fabbiano 1988).
The observational properties of such circumnuclear X-ray sources 
are summarized as follows (e.g., CM 99);
1) ROSAT X-ray luminosities, $L_{\rm X}$(0.2--2.4 keV)
$\sim 10^{37}$ -- $10^{40}$ ergs s$^{-1}$ with a mean of $3\times 10^{39}$
ergs s$^{-1}$; note that some of them are also detected
in the hard X ray and thus the total X-ray luminosities 
are higher by about one order of magnitude than the above values, 
2) the mean displacement between the location of the 
compact X-ray source and the optical photometric center of the galaxy
is $\sim$ 390 pc, and 3) they are common; the detection rate of such 
compact X-ray sources is $\gtsim$ 50\%.

Possible origins of these sources are; 1) black hole X-ray binaries,
2) low-luminosity AGN, 3) young X-ray luminous supernovae, or
4) a new X-ray population (e.g., CM99). Since the Eddington
luminosity of a 1.4 $\MO$ accreting neutron star is 
$10^{38.3}$ ergs s$^{-1}$, the luminous compact X-ray sources
cannot be explained by a single accreting neutron star.
If they are black hole X-ray binaries, their masses are
estimated to be $M_\bullet \sim$ 10$^2$ -- 10$^4 \MO$;
i.e., intermediate-mass black holes (hereafter IMBHs).
This urges us to consider how such IMBHs can be formed in 
circumnuclear regions of galaxies.

One of ideas may be that they are formed 
through the collapse of massive clouds (see Rees 1978).
In typical disk galaxies, gas clouds are present as a form of 
giant molecular gas clouds; hereafter GMCs.
A typical GMC has the following characteristics (e.g., Blitz 1993);
1) the mass, $M_{\rm GMC} \simeq$ (1 -- 2) $\times 10^5 \MO$,
2) the mean diameter, $D_{\rm GMC} \simeq 45$ pc,
3) the volume, $V_{\rm GMC} \simeq 9.6 \times 10^4$ pc$^3$,
4) the average density of H$_2$, $\overline{n}_{\rm H_2} \simeq 50$ cm$^{-3}$,
and 5) the mean column density of H$_2$, 
$\overline{N}_{\rm H_2} \simeq$ (3 -- 6) $\times 10^{21}$ cm$^{-2}$.
Although the typical kinetic temperature is $T_{\rm kin} \sim 10$ K,
the sound speed, $c_{\rm s}$,
is generally higher than the velocity dispersion
derived from the kinetic temperature because the turbulence
plays an important role in GMCs; therefore, $c_{\rm s} \sim 1$ km s$^{-1}$.
If a GMC is perturbed gravitationally by some physical mechanism,
it is likely that it begins to experience fragmentation,
resulting in the formation of sub-giant molecular clouds.
Such fragmentation could occur preferentially in denser parts of the GMC.
Since a typical H$_2$ density and a typical kinetic temperature 
in such dense parts are $n_{\rm H_2} \sim 10^4$ cm$^{-3}$ and
$T_{\rm kin} \sim$ 10 K, respectively, we obtain a Jeans' mass of
$m_{\rm J} \sim \lambda_{\rm J}^3 \rho_{\rm H_2} \sim 20 \MO$ for
$\lambda_{\rm J} = c_{\rm s} (\pi / G \rho_{\rm H_2})^{1/2}
\sim 0.36 ~ (c_{\rm s}/{\rm 0.3 ~ km ~ s^{-1}})$ pc and
$\rho_{\rm H_2} = n_{\rm H_2}  m_{\rm H_2}$
where $m_{\rm H_2}$ is the mass of a hydrogen molecule (Jeans 1929).
Therefore, IMBHs may not be formed directly from such massive gas clouds.

Instead, in this Letter, we propose that an IMBH is formed by 
continual merging 
of compact remnants (i.e., neutron stars and black holes)
left from massive stars in a giant H {\sc ii} region. The reason for this
is that giant H {\sc ii} regions are often observed in circumnuclear 
regions of disk galaxies (e.g., Kennicutt, Keel, \& Blaha 1989
and references therein; see for the formation mechanism of such
circumnuclear giant H {\sc ii} regions, Elmegreen 1994).
We also discuss some other proposed mechanisms responsible for
the formation of IMBHs. 

\section {Proposed Model}

\subsection{Dynamical Relaxation of Massive-Star Compact Remnants}

It has been considered that 
one plausible mechanism for the formation of
supermassive black holes (SMBHs) in galactic
nuclei is to pile up compact remnants of massive stars born in nuclear regions
(e.g., Spitzer \& Saslaw 1966; Spitzer \& Stone 1967, Spitzer 1969;
Saslaw 1973; Begelman \& Rees 1978; Weedman 1983; Norman \& Scoville 1988;
Quinlan \& Shapiro 1989, 1990; Lee 1995; Taniguchi, Ikeuchi, \& Shioya 1999).
On the analogy of this mechanism, we investigate the dynamical evolution of
compact remnants left from massive stars born in the core of a giant H {\sc ii}
region in circumnuclear regions of galaxies.

Typical masses of IMBHs lie in the range between 
$\sim 10^2 \MO$ and $\sim 10^4 \MO$ (CM99).
Therefore, we investigate a possibility that an IMBH with  
mass of $10^3 \MO$ can be formed from mergers of compact remnants
left in the core of a giant H {\sc ii} region. 
We consider a case that $N$ massive stars with masses higher
than 8 $\MO$ are formed within a sphere with a radius of $r_{\rm cl}$.
Here it is noted that all such massive stars yield a compact remnant
after the supernova explosion while the core of 
stars with masses less massive than 8 $\MO$ evolves into a white dwarf.
It is known that the mass of a neutron star
is $\approx 1.4 \MO$ while that of
a stellar-size black hole is from a few $\MO$ to  $\approx 10
\MO$; e.g., Brown \& Bethe (1994) suggested that main-sequence stars
with $18 \MO < m_* < 25 \MO$ evolve to low-mass black holes
($m_\bullet \simeq 1.5 \MO$) while stars with $m_* > 25 \MO$
evolve to high-mass black holes ($m_\bullet \gtsim 10 \MO$).
However, the mass function for giant H {\sc ii} regions has not yet
been well known. Therefore, in the above estimate,
we have assumed for simplicity that all the compact
remnants are black holes with a mass of 2 $\MO$ and thus
500 seed compact remnants are necessary to
make an IMBH with a mass of $10^3 \MO$.

All the massive stars die within a timescale of $\sim 10^7$ years 
and then a cluster of compact remnants will result in.
The candidates of IMBHs are observed in circumnuclear regions
of galaxies (i.e.,  a typical radial distance is
$R \simeq $ a few 100 pc). At such a distance, the rotation 
curve of galaxies may be described by the rigid rotation (e.g., Rubin et al.
1985; cf. Sofue 1997); we assume that the rigid rotation continues 
up to a radius of $R$ = 1 kpc and reaches the maximum rotation velocity of 
$v_{\rm rot}$ = 200 km s$^{-1}$.
If a giant H {\sc ii} region is formed at $R$ = 100 pc,
the cluster of massive stars (and thus the cluster of compact 
remnants too) is influenced by the mass contained within $R$ = 100 pc. 
This mass is estimated to be $M \sim R v_{\rm rot}^2 / G
\sim 9 \times 10^6 R_{100}^3 \MO$ where
$R_{100}$ is the radial distance
in units of 100 pc; note that $v_{\rm rot} = 20 R_{100}$ km s$^{-1}$.
The radius of the remnant cluster can be estimated as the tidal radius,

\begin{equation}
r_{\rm cl} \sim r_{\rm tidal} \sim R [N m_\bullet/(2M)]^{1/3}
\sim 3.8 N_{500}^{1/3} m_{\bullet, 2}^{1/3} ~~ {\rm pc}
\end{equation}
where $N_{500}$ is the number of compact remnants in units of 500 
and $m_{\bullet, 2}$ is the mass of a compact remnant in units of
2 $\MO$.  In this case, the cluster can be relaxed dynamically 
with a timescale of

\begin{equation}
\tau_{\rm dyn} \sim N^{1/2} G^{-1/2} r_{\rm cl}^{3/2} m_\bullet^{-1/2}
\sim 1.8 \times 10^9 ~~ {\rm y}.
\end{equation}
Since this timescale is shorter than the age of galaxies,
this mechanism may be responsible for the formation of IMBHs in
circumnuclear regions of galaxies.

It is noted that the dynamical situation for a cluster of compact
remnants considered here is completely different from both that
for a cluster of compact remnants in globular clusters (e.g.,
Kulkarni, Hut, \& McMillan 1993) and that for a cluster of
compact remnants in the nuclear region
of galaxies (e.g., Weedman 1983; Lee 1995 and references therein)
because low-mass stars dominate the cluster gravitational potential
for the latter two cases. On the other hand, it is expected that
massive stars may be preferentially formed in the core. Since supernovae
could blow the gas resided originally in the core region, compact
remnants themselves may dominate the cluster gravitational potential.
Furthermore, lifetimes of OB stars are less than $\sim 10^7$ years,
being much shorter than the dynamical relaxation timescale 
estimated above. Therefore,
the star cluster evolves to a cluster of compact remnants quickly.
Some black holes may be ejected from the cluster by the 
dynamical effect during the formation of black hole binaries through
the three-body encounters (Spitzer 1987; Kulkarni et al. 1993).
However, this effect is negligible if $N \gtsim 100$ 
(e.g., Binney \& Tremaine 1987); this is indeed the case discussed here.

Then we investigate what kind of massive star formation is necessary.
We estimate an average number density of gas,
$n_{\rm H} = M_{\rm gas} / [(4 \pi/3) r_{\rm cl}^3 m_{\rm H}]
\simeq 1 \times 10^4$ cm$^{-3}$ where 
$M_{\rm gas} = N m_* \eta_{\rm SF, 0.1}^{-1} = 5 \times 10^4 \MO$ 
given $m_* = 10 \MO$ and 
$\eta_{\rm SF, 0.1}$ is the star formation efficiency 
in units of 0.1; i.e., 10\% of the gas in the cloud is
used to form massive stars. 
This density is comparable to those in typical
dense cores in star forming regions (e.g., $n_{\rm H}
\sim 10^4$ cm$^{-3}$; e.g., Lada, Strom, \& Myers 1993). 
Note that the gas mass lies in the observed range of $M_{\rm GMC}$. 
The mass volume density of massive stars is estimated as 
$ N m_* / [(4 \pi/3) r_{\rm cl}^3] \simeq 22 \MO$
pc$^{-3}$. This is lower by two orders of magnitude than
that of the compact star cluster R136a, whose 
half light radius is 1.7 pc, embedded in the
central region of the 30 Dor nebula
in Large Magellanic cloud; $5.5 \times 10^4 \MO$ pc$^{-3}$
(Hunter et al. 1995). 
Therefore, the star formation properties as well as the star cluster 
properties in our model are not unusual.

Finally we mention which type of galaxies tends to have 
more massive IMBHs. Using equations (1) and (2) together with
the relation $M \sim R v_{\rm rot}^2 / G$, we obtain

\begin{equation}
\tau_{\rm dyn} \sim {N \over 2} ~ {R \over v_{\rm rot}} 
\sim {N \over 2} ~ {\tau_{\rm rot} \over {2 \pi}},
\end{equation}
where $\tau_{\rm rot}$ is the rotation period. If the rigid rotation
is achieved up to a radius of $R =$ 1 kpc as observed,
the dynamical timescale is independent of $R$ if $R \leq 1$ kpc; 
i.e., any cluster of compact remnants left from the core of a giant
H {\sc ii} region within $R \leq$ 1 kpc could evolve to an IMBH.
Since any disk galaxy have been able to form such IMBHs during its
lifetime, i.e., $\sim 10^{10}$ years, the total number of IMBHs formed
in its lifetime is estimated to be 

\begin{equation}
N \sim 4 \pi \times \left({{\tau_{\rm dyn}} \over {\tau_{\rm rot}}}\right) 
\sim 4 \times 10^3 \left({{\tau_{\rm dyn}} \over {10^{10} ~ {\rm y}}}\right)
 \left({{\tau_{\rm rot}} \over {3.1 \times 10^{7} ~ {\rm y}}}\right)^{-1}.
\end{equation}

More importantly, the relation given in equation (3)
implies that galaxies with slower rigid-rotation velocities 
(i.e., late-type spirals) tend to have less massive IMBHs
while those with higher rigid-rotation velocities (i.e.,
early-type spirals) tend to have more massive IMBHs. 
Since the survey conducted by CM99 is biased to late-type spirals,
we are unable to make this observational test. X-ray surveys will be 
recommended for a sample of early-type spirals to test whether or not
this prediction is consistent with observation. 

\subsection{Intermediate-mass Black Holes Supplied 
from Satellite Galaxies}

We have shown that an IMBH can be made through the merger of
compact remnants of massive stars born in a giant H {\sc ii} region.
Since any galaxies have satellite galaxies (e.g., Zaritsky et al. 1997) 
and giant H {\sc ii} regions could be made in some gas-rich satellites
(e.g., 30 Dor in LMC),  it seems important to
investigate a possibility that IMBHs can be supplied
by minor mergers with satellite galaxies having IMBHs.
M 32, one of the satellite galaxies of M 31, may have a 
supermassive black hole with a mass of $\sim 10^6 \MO$
(Dressler \& Richstone 1988; see for a review Kormendy et al.
1998). At present, there is no observational evidence
for the presence of IMBHs in any satellite galaxies.
However, if some satellite galaxies had IMBHs in their
nuclei, it could be possible that minor mergers with such satellites
are responsible for the presence of IMBHs in circumnuclear regions
of the host galaxies.

We estimate the frequency of occurrence of compact X-ray sources
if all of them are attributed to the IMBHs supplied by minor mergers with
nucleated satellite galaxies (see Taniguchi \& Wada 1996).
Tremaine (1981) estimated that every galaxy would experience minor mergers with
its satellite galaxies several times. Since a typical galaxy may have several
satellite galaxies, the probability of merger
for a satellite galaxy may be estimated to be $f_{\rm merger} \simeq 0.5$; i.e.,
half of the satellite galaxies have already merged to a host galaxy, while the rest
are still orbiting. Another important value is the number of nucleated
satellite galaxies. For example,
M31 has two nucleated satellites (M32 and NGC 205), and a field S0 galaxy
NGC 3115 has a nucleated dwarf (van den Bergh 1986).
Although there has been no systematic search for nucleated satellite galaxies,
it is likely that  every galaxy has (or had) a few nucleated satellites.
Therefore, for simplicity we assume $n_{\rm sat} = 2$.
If we assume that the typical lifetime of the X-ray sources
is $\tau_{\rm active} \simeq 10^8$ years, we obtain an 
expected frequency,

\begin{equation}
P_{\rm IMBH} \simeq f_{\rm merger} ~ n_{\rm sat} ~ \tau_{\rm active} ~
\tau_{\rm Hubble}^{-1} \sim 0.01 n_{\rm sat, 2} \tau_{\rm active, 8},
\end{equation}
where $\tau_{\rm Hubble}$ is the Hubble time, $\sim 10^{10}$ years,
$n_{\rm sat, 2}$ is the number of nucleated satellites in units of 2,
$\tau_{\rm active, 8}$ is the duration of the active phase
in units of $10^8$ years. Hence,
if minor mergers with nucleated satellites are responsible for
the IMBHs in the circumnuclear regions of their host galaxies, 
it is statistically expected that hard X-ray sources
are found in about 1 \% of field disk galaxies.
This is significantly smaller than the observed frequency, $\gtsim$
50\% (CM99).

However, it seems possible that IMBHs can be formed through 
the dynamical relaxation of cores of giant H {\sc ii} regions
in gas-rich satellite galaxies and they travel 
to circumnuclear regions of their host galaxies by minor mergers.
This case gives another expected frequency, 

\begin{equation}
P_{\rm IMBH} \simeq f_{\rm merger} ~ n_{\rm sat} ~ n_{\rm IMBH} ~ 
\tau_{\rm active} ~ \tau_{\rm Hubble}^{-1}
\end{equation}
where $n_{\rm IMBH}$ is the number of IMBH in each satellite galaxy.
Since the maximum number of IMBHs can be estimated as 
$n_{\rm IMBH} \sim \tau_{\rm Hubble} / \tau_{\rm dyn}$, we obtain
 
\begin{equation}
P_{\rm IMBH} \simeq f_{\rm merger} ~ n_{\rm sat} ~  
\tau_{\rm active} ~ \tau_{\rm dyn}^{-1} \sim
0.1 n_{\rm sat, 2} \tau_{\rm active, 8} \tau_{\rm dyn, 9}^{-1}
\end{equation}
where $\tau_{\rm dyn, 9}$ is the dynamical timescale in units of
$10^9$ years. Note that $n_{\rm sat}$ is not the number of nucleated
satellites but that of gas-rich satellites. However, for simplicity,
we adopt $n_{\rm sat} = 2$ even in this estimate.

In summary, the frequency of occurrence of IMBHs supplied from
satellite galaxies lies in a range between 0.01 and 0.1.
Therefore, we suggest that 
minor mergers may not explain the observed  higher frequency of IMBHs
unless $n_{\rm sat} > 10$.

\section{Alternative Mechanisms}

We have shown that continual merging of compact remnants left from
massive stars in a giant H {\sc ii} region formed in circumnuclear
regions of galaxies may be responsible for the formation of an IMBH
intermediate-mass black hole within a timescale of $\sim 10^9$ years.
A necessary condition is that several hundreds of massive
stars are formed in a compact region with a radius of a few pc.
In this section, we discuss some alternative mechanisms responsible
for the formation of IMBHs.

\subsection{Bondi-type gas accretion onto a seed black hole}

Compact remnants left from massive stars formed in the initial 
star formation in a galaxy have been surviving for the galaxy age,
i.e., $\tau_{\rm age} \sim 10^{10}$ years. 
They have been experiencing a number of encounters with gas clouds
in the galaxy. Therefore, classical Bondi-type (Bondi 1952) gas accretion
is the most probable accretion process for them (Yoshii 1981).
This gas accretion rate is given by
$\dot M_{\rm Bondi} = 2~\pi ~m_{\rm H} ~ n_{\rm H} ~ r_{\rm a}^2 ~ v_{\rm e}$,
where $m_{\rm H}$, $n_{\rm H}$, $r_{\rm a}$, and $v_{\rm e}$ are 
the mass of a hydrogen atom, the number density of the hydrogen atom,
the accretion radius defined as $r_{\rm a} = G M_\bullet v_{\rm e}^{-2}$ 
($M_\bullet$ is the mass of the seed compact remnant), 
and the effective relative velocity 
between the seed black hole and the ambient gas, respectively.

Since the Bondi-type gas accretion is most important for 
the low-velocity encounter, we adopt  $v_{\rm e} = 1$ km s$^{-1}$
in order to estimate the maximum accretion rate
(Yoshii 1981) although $v_{\rm e}$ is realistically much higher
than this value.
Adopting an average gas density of the interstellar medium,
$n_{\rm H} \sim 1$ cm$^{-3}$, we obtain the total accreting mass
due to the Bondi-type accretion,

\begin{equation}
M_{\rm acc, Bondi} = {\dot M}_{\rm Bondi} ~ \tau_{\rm age} 
\sim 3 \times 10^{-2} M_{\bullet, 1}^2
n_{\rm H, 1} v_{\rm e, 1}^{-3} \tau_{\rm age, 10} ~ \MO
\end{equation}
where $M_{\bullet, 1}$ is the mass of  the seed compact remnant
in units of $1 \MO$, 
$n_{\rm H, 1}$ is the average gas density in units of $1$ cm$^{-3}$,
$v_{\rm e, 1}$ is the orbital velocity with respect to the ambient gas
in units of 1 km s$^{-1}$, and $\tau_{\rm age, 10}$ is the age of 
the galaxy in units of $10^{10}$ years.
Even if a black hole with a mass of 10 $\MO$ was born 
$10^{10}$ years ago in the galaxy (see for the formation of 
black holes with $m_\bullet \simeq 10 \MO$, Brown \& Bethe 1994), 
its accreting mass amounts only to a few $\MO$.
It is expected that some black holes might encounter dense molecular
clouds. However, the crossing time is as short as
$\tau_{\rm cross} \sim D_{\rm GMC, 45} / v_{\rm e, 1} \sim
4.5 \times 10^7$ years where $D_{\rm GMC, 45}$ is the mean
diameter of GMCs in units of 45 pc. 
It is also noted that the age of a GMC may range between $\sim 10^5$
years (the molecule formation time) and several times 10$^8$ years
(e.g., Elmegreen 1985). Therefore, in the case of an encounter
between a black hole and a GMC,  we should adopt $\tau_{\rm age}
\sim 10^8$ years in equation (8).  Therefore,
the mass growth is negligibly small
even if $\overline{n}_{\rm H} \sim 50$ cm$^{-3}$.
However, if $v_{\rm e} \ll 1$ km s$^{-1}$, an IMBH could be formed
through this accretion process.
Since it seems unlikely that such a very slow encounter occurs
frequently, it is suggested that the Bondi-type gas accretion
onto the seed black hole may not be responsible for the formation of
an IMBH within the age of the galaxy.

\subsection{Disk-type gas accretion onto a seed black hole}

Next we consider a case that a compact remnant is a black hole
with a mass of 1 $M_\bullet$ and the disk-type gas accretion has been
occurring at the Eddington accretion rate, 
${\dot M}_{\rm Disk} \simeq 2.2 \times 10^{-8} \eta_{\rm acc, 0.1} 
M_{\bullet, 1} ~ \MO ~ {\rm y}^{-1}$
where $\eta_{\rm acc}$ is the conversion efficiency from
the gravitational energy to the radiation in units of 0.1 (e.g., Rees 1984).
The seed black hole with a mass of 1 $\MO$ can increase in mass and its
mass is given by 

\begin{equation}
M_\bullet(t_8)  = 
M_{\bullet, 1} e^{2.2 \eta_{\rm acc, 0.1} t_8} ~~ \MO
\end{equation}
where $t_8$ is the duration of the gas accretion in units of
$10^8$ years.
One obtains $M_\bullet = 10^2 \MO$ at $t_8 = 2.09$,
$M_\bullet = 10^3 \MO$ at $t_8 = 3.14$, and
$M_\bullet = 10^4 \MO$ at $t_8 = 4.19$.
Therefore, it seems that IMBHs could be easily formed by this mechanism.

However, there is a serious problem in this model.
The star formation rate may be higher in earlier phases of
galaxy evolution. Even if we assume that massive stars
have been made continuously at a constant rate,
the number of compact remnants made during the last $10^9$ years
is estimated to be $\sim 10^7$ for a typical galaxy  with mass of 
$10^{12} \MO$ in which stars were formed with the Salpeter mass
function (see Taniguchi et al. 1999).
If all these compact remnants have been
experiencing the disk accretion at the Eddington rate,
we would observe a lot of IMBHs in the galaxy if they have not
yet merged into one SMBH. 
If they were already merged into one SMBH, there would be
a very supermassive black hole with mass of $M_\bullet 
\sim 10^7 \times 10^4 \MO \sim 10^{11} \MO$ in some galaxies.
Therefore, the idea described here may not be a 
dominant mechanism responsible for the formation of IMBHs.

\par
\vspace{1pc}\par
We would like to thank an anonymous referee for his/her many
useful comments and suggestions which improved this paper
significantly.
This work was supported in part by the Ministry of Education, Science,
Sports and Culture in Japan under Grant Nos. 10044052, and 10304013.

\clearpage
\section*{References}

\re 
Blitz L.\ 1993, in Protostars and Planets III, 
ed E.H. Levy, J.I. Lunine (The University of Arizona 
Press, Tucson), p125
\re 
Begelman M.C., Rees M.J.\ 1978, MNRAS, 185, 847
\re 
Binney J., Tremaine S.\ 1987, Galactic Dynamics 
(Princeton University Press, Princeton), 492
\re 
Bondi H.\ 1952, MNRAS, 112, 195
\re 
Brown G.E., Bethe H.A.\ 1994, ApJ, 423, 659
\re 
Colbert E.J.M., Mushotzky R.F.\ 1999, ApJ, 519, 89
\re 
Colbert E.J.M., Petre R., Schlegel E.M., Ryder S.D.\ 
1995, ApJ, 446, 177
\re 
Dressler A., Richstone D.O.\ 1988, ApJ, 324, 701
\re
Elmegreen B.G.\ 1985, in Protostars and Planets, edited by D. C.
Black, and  M. S. Matthews (The University of Arizona Press), 33
\re
Elmegreen B.G.\ 1994, ApJ, 425, L73
\re
Fabbiano G.\ 1988, ARA\&A, 27, 87
\re
Fabbiano G., Trinchieri G.\ 1987, ApJ, 315, 46
\re
Hunter D.A., Shaya E.J., Holtzman J.A., Light R.M., 
O'Neil E.J., Lynds R.\ 1995, ApJ, 448, 179
\re
Jeans J.H. 1929, Astronomy and Cosmogony, 2nd ed. 
Cambridge, Eng.:  (Cambridge University Press, Cambridge)
\re
Kennicutt R.C.\ Jr, Keel W.C., Blaha C.A.\ 1989, 
AJ, 97, 1022
\re
Kohmura Y., et al.\ 1994, PASJ, 46, L157
\re
Kormendy J., Bender R., Evans A.S., Richstone D.\ 1998, 
AJ, 115, 1823
\re
Kulkarni S.R., Hut P., McMillan S.\ 1993, Nature, 364, 421
\re
Lada E.A., Strom K.M., Myers P.C.\ 1993, 
in Protostars and Planets III, ed E.H. Levy, J.I. Lunine 
(The University of Arizona Press, Tucson), p245
\re
Lee H.M.\ 1995, MNRAS, 272, 605
\re
Makishima K., et al. 2000, ApJ, submitted (astro-ph/0001009)
\re 
Matsumoto H., Tsuru T.G.\ 1999, PASJ, 51, 321
\re
Mushotzky R. F., Done C., Pounds K. A.\ 1993, ARA\&A, 31, 717
\re
Norman C., Scoville, N.\ 1988, ApJ, 332, 124
\re
Okada K., Dotani T., Makishima K., Mitsuda K., Mihara T.\ 
1998, PASJ, 50, 25
\re
Petre R., Okada K., Mihara T., Makishima K., Colbert E.J.M.\ 
1994, PASJ, 46, L115
\re
Ptak A., Griffiths R.\ 1999, ApJ, 517, L85
\re
Rees M.J.\ 1978, Observatory, 98, 210
\re
Rees M.J.\ 1984, ARA\&A,  22, 471
\re
Reynolds C.S., Loan A.J., Fabian A.C., Makishima K., 
Brandt W.N., Mizuno T.\ 1997, MNRAS, 286, 349
\re
Rubin V.C., Burstein D., Ford W.K.\ Jr, Thonnard N.\ 
1985, ApJ, 289, 81
\re
Quinlan G.D., Shapiro S.L.\ 1989, ApJ, 343, 725
\re 
Quinlan G.D., Shapiro S.L.\ 1990, ApJ, 356, 483
\re
Salpeter, E.E. 1955, ApJ, 121, 161
\re
Saslaw W.C.\ 1973, PASP, 85, 5
\re
Sofue Y.\ 1997, PASJ, 49 17
\re
Spitzer L.\ 1969, ApJ, 158, L139
\re
Spitzer L.\ 1987, Dynamical Evolution of Globular Clusters 
(Princeton University Press, Princeton), ch3
\re
Spitzer L., Saslaw W.C.\ 1966, ApJ, 143, 400
\re
Spitzer L., Stone, M.E.\ 1967, ApJ, 147, 519
\re
Stevens I.R., Strickland D.K., Wills K.A. MNRAS, 308, L23
\re
Takano M., Mitsuda K., Fukazawa Y., Nagase F. 1994, 
ApJ, 436, L47
\re
Taniguchi Y., Ikeuchi S., Shioya Y.\ 1999, ApJ, 514, L9
\re
Taniguchi Y., Wada K.\ 1996, ApJ, 469, 581
\re
Tremaine S. 1981, in The Structure and Evolution of Normal Galaxies, 
ed S.M. Fall, D. Lynden-Bell (Cambridge University
Press, Cambridge), p67
\re
van den Bergh S.\ 1986, AJ, 91, 271
\re
Weedman D.W.\ 1983, ApJ, 266, 479
\re
Yoshii Y.\ 1981, A\&A, 97, 280
\re
Zaritsky D., Smith R., Frenk C., White S.D.M. 1997,
ApJ, 478, 39

\end{document}